\documentstyle[12pt]{article}
\input epsf

\begin{document}
\begin{titlepage}
\thispagestyle{empty}
\title{
\vspace*{-1cm}
\begin{flushright}
{\small CPHT S 040.0400}
\end{flushright}
\vspace{2.0cm}
Reinterpretation of Effective Chiral Lagrangian }
\vspace{4.0cm}
\author{Tran N. Truong \\
\small \em Centre de Physique Th\'eorique, 
{\footnote {unit\'e propre 014 du
CNRS}}\\ 
\small \em Ecole Polytechnique \\
\small \em F91128 Palaiseau, France}

\date{April 2000}

\maketitle

\begin{abstract}
Effective Tree Chiral Lagrangian is interpreted as a power series expansion of the
kinematical variables. In the
presence of the strong interaction this expansion is valid below the unitarity cut, hence in
the unphysical region. Consequences of this reinterpretation of the Chiral Lagrangian
are analysed for the relation  between $K-\pi$ and $K-2\pi$ transitions.

\end{abstract}
\end{titlepage}

There has been recent interest how to handle the problem of the strong final state pion pion 
interaction in $K_s \rightarrow 2\pi$
decay. In particular how to determine the off-shell $K-\pi$ transition using the input of the
measured $K \rightarrow 2\pi$ rate. This last reaction is quite difficult to calculate by
the technique of the lattice gauge theory because of the strong  $\pi \pi$ final state
interaction and of other reasons. The
$K-\pi$ transition should be easier to calculate using the lattice technique and could
provide the answer to the origine of the
$\Delta I=1/2$ problem \cite{truong1}. 

In a related problem which is now of central interest, is to understand the experimental
ratio $\epsilon^{\prime}/\epsilon$ of the CP violation problem. Eventually, using the
lattice  theory, one should be able to calculate numerically this CP violation effect
 by reducing them to the $K-\pi$ problem. We note that there is a recent  calculation of this
ratio  taking into account of the strong final state interaction using the
technique discussed previously
\cite{pallante}.

A possible solution for the first problem was given a long time ago \cite{truong1}. It was
  based on a reinterpretation of the Effective Chiral Lagrangian (CL) at the tree level 
which has not been clear to all readers. This has led to some questions raised by a
number of authors
\cite{neubert} and more recently by Buras et
al.\cite{buras}. We wish to clarify in this note some points raised by these authors. It is
shown here that interpreting the tree CL  of the process $K \rightarrow 2\pi$
amplitude as a power series expansion in  momentum together with requiring the zero of the
matrix element as demanded by the Cabibbo and Gell-Mann theorem \cite{cab}, assure that the final
result is  model independent. Our line of approach has been used in a series of articles
\cite{truong2, truong3, truong4, truong6, truong5}.

In this note the problem of the final state interaction is reexamined.  The main idea is that,
just the same as in our previous work,  the tree CL is an explicit manifestation of the current
algebra soft pion theorems. It was first invented to avoid some complicated and cumbersome
manipulations of the current algebra technique \cite{adler}. It is crucial to note that the
current algebra soft pion theorems are still valid  in the presence of  the hadronic initial or
final state interactions. The current algebra relation between the form factors of
$K \rightarrow \pi\pi e\nu$ and
$K \rightarrow \pi e \nu$ is such an example \cite{truong3, truong5}. It is still valid in the
presence of  the strong interaction between $K\pi$ and
$\pi\pi$...

As  was previously suggested, the tree CL, should bear this important property
i.e it should be valid even in the presence of the strong interactions among the hadrons
involved. At first sight this cannot be done because CL only gives a power series expansion
in the invariant variables $s, t, u ...$ of the matrix element which must be real in the
physical region while the strong final state interaction should make them complex. There is,
however, a region where the power  series expansion of the matrix element is valid, namely
outside the cut
 in the unphysical region.  We shall make use of the CL to
give relations among different processes just the same as the current algebra technique. The tree
CL  can therefore be considered as low energy theorems in the unphysical region for
our purpose.  It is not difficult to analytically continue these low energy theorems to the
physical region by using the technique of the integral equations of the Omnes-Muskhelishvili type
(OM) 
\cite{omnes,mus}, or the inverse amplitude method \cite{truong5} etc.

 The effective Lagrangian for the $\Delta I=1/2$ $K\rightarrow 2\pi$ is given by \cite{cronin}: 
\begin{equation}
M(K_S(k)\rightarrow \pi^+(p)+\pi^-(q) )= \frac{1}{\sqrt{2}}Cf_\pi(2k^2-p^2-q^2)
\label{eq:k2pi}
\end{equation}
and
\begin{equation}
 M(K_L\rightarrow\pi^0)=-iC\sqrt{2}f_\pi^2q(\pi).k(K)
\label{eq:kpi}
\end{equation}
where $f_\pi=93 MeV$ and is the pion decay constant. Notice the constant C are common to
both equations which the manifestation of the CA soft pion theorem relating the off-shell
$K-\pi$ to $K\rightarrow 2\pi$ amplitude. 

 Let us consider the Eq. (\ref{eq:k2pi}) when both pions are
on their mass shell and the Kaon off its mass shell. The usual interpretation of this 
equation is simply the result of the effective tree Lagrangian or the result with the
strong final state $\pi\pi$ interaction switched off. 
To take into account of the final state $\pi \pi$ interaction, Chiral Perturbation Theory
(ChPT) could be used. Because of the presence of the undetermined counterterms and of the
violation of the unitarity inherited in the perturbative schemes, the phase theorem is no
longer satisfied in this scheme  and hence this approach is not useful for our
purpose.

 The crucial point is to reinterprete
Eq. (\ref{eq:k2pi}) as the result of the first two terms of a power series expansion in $s=k^2$
variable of the 
 matrix elements   $K
\rightarrow 2\pi$ with the presence $\pi\pi$, $4\pi$, $K \pi$ ... interactions. This expansion
is only valid in the unphysical region. Let us denote this matrix element with the two pions on
their mass shell as
$A(s)$. It is assumed that
$A(s)$
 is analytic in the cut plane with a cut from
$4m_\pi^2$ to
$\infty$. In reality, $A(s)$ 
 is a product of two functions, the self energy operators of the Kaon  and the $K-2\pi$
vertex. (We shall neglect in the following the Kaon self energy operators due to its higher
threshod $K2\pi$...).

Let us first discuss the solution of this problem from a more general viewpoint. In our
non-perturbative approach, the effective tree CL  represents low
energy theorems with strong (final state) interaction taken into account. Below the cut  their
contribution can be represented by a polynomial in $s$ of  degree $n$ and with real coefficients.
Without loss of generality this polynomial can also be rewritten as as a polynomial in $(s-s_0)$
variable where
$s_0$ is in the unphysical region which will be taken on the real $s$ axis below the branch point
$4m_\pi^2$. 

Assuming that $A(s)$ is polynomially bounded and that $A(s)s^{-(n+1)} \rightarrow 0$ as $s
\rightarrow \infty$, $n>0$, we  can write the following dispersion relation
\begin{equation}
A(s,s_0)=a_0+a_1(s-s_0)+...a_{n}(s-s_0)^{n}+ \frac{(s-s_0)^{n+1}}{\pi}\int_{4m_\pi^2}^\infty
\frac{ImA(z)dz}{(z-s_0)^{n+1}(z-s-i\epsilon)}
\label{eq:dr} 
\end{equation}

Around $s=s_0$, the dispersion integral is of the order
$(s-s_0)^{n+1}$ and can be neglected, hence the low energy theorem is recovered. Needless to
say, $a_n$ are, apart from a factorial factor $n!$, the derivatives of $A(s)$ evaluated at
$s=s_0$.

The mathematical problem is now  clear: Find the solution of the integral
equation of the OM type for $A(s,s_0)$ with  its imaginary part given by the elastic
unitarity:
\begin{equation}
ImA(s)= A(s,s_0) e^{-i\delta (s)} sin \delta (s) \label{eq:imo}
\end{equation}
 with the boundary conditions around $s_0$ given by Eq. (\ref{eq:dr}) and where $\delta$ is the
S-wave $I=0$ $\pi\pi$ phase shifts.

To solve this integral equation Eq. (\ref{eq:dr}) which is of OM type
\cite{omnes,mus},  let us define the function 
$\Omega(s,s_0)$  normalized to unity for convenience at $s=s_0$:
\begin{equation}
\Omega(s,s_0)= \exp(\frac{s-s_0}{\pi} \int_{4m_\pi^2}^\infty
\frac{\delta(z)dz}{(z-s_0)(z-s-i\epsilon)})
\label{eq:omn}
\end{equation}
 The solution for our integral equation is:
\begin{equation}
A(s,s_0) = P_n(s)\Omega(s,s_0) \label{eq:solo}
\end{equation}
where $P_n(s)$ is a  polynomial in $s$ of order n with real coefficients. They can be determined
by expanding the function $\Omega(s,s_0)$ in a power series in
$(s-s_0)$,
  and compare Eq. (\ref{eq:solo}) with
Eq. (\ref{eq:dr}). The expansion in Taylor's series  is possible because  
$\Omega(s,s_0)$  is an analytic function with a cut from $4m_\pi^2$ to $\infty$.

In the special case where only two terms in the series are known such as the case of the CL
given by Eq. (\ref{eq:k2pi}), the solution of our integral equation is given  by:

\begin{equation}
A(s)= \{a_0+ (s-s_0)(a_1-a_0\Omega^{'}(s_0,s_0)) \}\Omega(s,s_0) \label{eq:sol}
\end{equation}
where $\Omega^{'}$ denotes the first derivative of $\Omega(s,s_0)$ evaluated at $s_0$
and is given by:
\begin{equation}
\Omega{'}(s_0,s_0)=\frac{1}{\pi} \int_{4m_\pi^2}^\infty \frac{\delta(z)dz}{(z-s_0)^2}
\label{eq:der}
\end{equation}
The presence of the term $\Omega^{'}(s_0,s_0)$ is to ensure the boundary condition for
$A(s,s_0)$ is  satisfied. 

 It is straightforward to generalise the solution of Eq. (\ref{eq:dr}) for other values of
$n$. For example when  $n=2$, the solution for the integral equation Eq.( \ref{eq:dr}) is
obtained by adding to the curly bracket on the righthand side of Eq. (\ref{eq:sol}) a 
term:
\begin{equation}
(s-s_0)^2(
a_2-a_1\Omega^{'}(s_0,s_0)-a_0\frac{\Omega^{''}(s_0,s_0)}{2}+a_0\Omega^{'2}(s_0,s_0))
\label{eq:n=2}
\end{equation}

One is tempted to write a simpler solution than that given by Eq. (\ref{eq:sol}) by
construcing for example:
\begin{equation}
A(s,s_0)=(a_0+a_1 (s-s_0))\Omega_2(s,s_0)
\end {equation}
where 
\begin{equation}
\Omega_2(s,s_0)= \exp(\frac{(s-s_0)^2}{\pi} \int_{4m_\pi^2}^\infty
\frac{\delta(z)dz}{(z-s_0)^2(z-s-i\epsilon)})
\label{eq:omn2}
\end{equation}
which satisfies the boundary conditions, but violates the condition on the
polynomially boundedness. This is so because by partial fraction, one can show
$\Omega_2(s,s_0)=\Omega(s,s_0)/\exp((s-s_0)\Omega{'}(s_0,s_0))$ which has an exponential
behavior. This result is totally expected because the dispersion relation for
$\log\Omega(s,s_0)$ obeys at most a once subtracted dispersion relation due to the
polynomial boundedness.

 As usual the
solution given by Eq. (\ref{eq:sol}) has the polynomial ambiguity because we can multiply the
RHS of Eq. (\ref{eq:sol}) by a polynomial factor $1+\sum_{n=2}^{N}c_n(s-s_0)^n$ with $N\geq 2 $.
We assume, in the following, there is no such ambiguity, or the zeros introduced by such an
ambiguity is sufficiently far away from the physical region of interest.

 For the $K \rightarrow \pi\pi$ problem, 
invoking the Cabibbo and Gell-Mann theorem \cite{cab} which requires the matrix element to vanish
in the SU(3) limit, one has 
$s_0=m_\pi^2$.

We now return to our problem. Let us rewrite Eq. (\ref{eq:k2pi}) with the two pion on their
mass shell as:
\begin{equation}
M(K_S(k)\rightarrow \pi^+ (p) +\pi^- (q) ) =
\sqrt{2}Cf_\pi (s-m_\pi^2)+...\label{eq:K2pi}
\end{equation}
This equation should be considered as the power  series around $s=m_\pi^2$ keeping only
the first 2 terms, namely the matrix element vanishes at $s=m_\pi^2$ and its derivative at
this point is known. The solution for the corresponding Omnes-Muskhelishvilli equation
\cite{omnes,mus} is therefore:
\begin{equation}
M(s)= \sqrt{2}Cf_\pi(s-m_\pi^2)\Omega(s,m_\pi^2) \label{eq:1}
\end{equation}
as can be seen using the result of Eq. (\ref{eq:sol}) and letting $a_0=0$ and $s_0=m_\pi^2$ as
required by Eq. (\ref{eq:K2pi}). The
condition on the position of the zero of the matrix element at
$s=m_\pi^2$ is
 a direct consequence of the Cabibbo-Gell-Mann theorem on the SU(3) symmetry of
the problem \cite{cab}.  Eq. (\ref{eq:1}) were derived earlier without giving explicitly  a
justification \cite{truong1}. 

  The physical value of the matrix element is obtained by setting $s=m_K^2$.

Using the
experimental rate for $K_s\rightarrow 2\pi$ and the S-wave,
$I=0$ phase shifts as given by the unitarized one loop ChPT which fit to the experimental
data \cite{truong1, truong5}, one obtains
$C=0.90.10^{-11} MeV^{-2}$. This is the result of the reference \cite{truong1}.

Our strategy to study the $K_s\rightarrow 2\pi$ and $K_L\rightarrow 3\pi$ is therefore to
calculate first the $K-\pi$ transition by lattice gauge theory or by some approximate schemes
\cite{all} and compare them with the value $C$ given above.

For other problems, such as the $\eta \rightarrow \pi^+ \pi^- \pi^0$, because there is no Cabibbo
and Gell-Mann theorem when the $\pi^0$ and the $\pi^-$ are soft, how can $s_0$ be determined ? An
approximate answer to this problem is to realise that the effective CL is a realisation of the
Current Algebra soft pion theorems
\cite{truong4, truong6}. By taking $\pi^0$ soft, using current algebra, this problem is reduced to
the
 matrix element $<\eta\mid v\mid\pi^+(q_1)\pi^-(q_2)>$ where
$v$ is a pseudo-scalar operator. This matrix element is similar to the $K \rightarrow
2\pi$ problem treated here. Taking one of the remaining two pions soft, the other pion on its mass
shell, this matrix element is related to the
$\eta-\pi$ mixing problem. Going through this process for the $\eta \rightarrow 3\pi$ problem,
one has 
  $s=(q_1+q_2)^2 \rightarrow m_\pi^2 =s_0$ when $q_1\rightarrow 0$. This point  sets the scale
to this problem in terms of the $\eta-\pi$ mixing
\cite{truong4}. The corresponding dispersion relation for this problem is similar to Eq.
(\ref{eq:dr}), except only a once subtraction at $s_0$ is needed, with $a_0$ given by the value of
the
$\eta-\pi$ mixing.

This value of $s_0$ is different from that given in the reference \cite{truong6}, namely $s_0=0$,
the chiral $SU(3)$ limit. The final results is insensitive to these two choices of $s_0$. They
differ from each other by only a few percents.

We have presented  here a reasonable method to determine $s_0$. This method is inspired by the
current algebra technique \cite{truong6}.

This discussion of the $\eta \rightarrow \pi^+ \pi^- \pi^0$ problem is also of interest for
calculating the matrix element of $(8_L,8_R)$ operators of the $\Delta I=3/2$ $K\rightarrow
 2\pi$ problem.

The following  discussion on the subtraction point $s_0$, based on the theory of analytic
function is of some interest. Because
$A(s,s_0)$ is an analytic function in the cut s-plane, its Taylor's series converges inside a
circle with a center at $s_0$ and of a radius
$\mid4m_\pi^2-s_0\mid$. The choice  of the number of terms in the series to achieve a given
accuracy depends on the physical situation and also on the choice of the point $s_0$.

 For example for
$s_0$ near to the branch point $s=4 m_\pi^2$, the radius of the convergence of the series is very
small and hence more terms are needed in the power series if the series converges at all. For
$s_0=4m_\pi^2$, the radius of the convergence is zero which is expected because all the
derivatives of $A(s)$ evaluated at this point become infinite due to the square root threshold
singularity  and hence the series diverges.

If the expansion point $s_0$ was taken far away from the origin and on the negative s axis, the
radius of the convergence of the series would be, in principle, larger but  we would have to take
more terms in the series in order to explore the boundary conditions near the origin which is the
chiral limit of the matrix element. 

As it is shown above, some physical input must be made to restrict the determination of $s_0$.

To see the sensitivity of our solution on the expansion point $s_0$  we
 pretend to ignore the Cabibbo-Gell Mann theorem and study the solution of
Eq.(\ref{eq:sol}) as a function of $s_0$. Let us take $s_0=0$ and $s_0=2m_\pi^2$. The former
yields the zero of the $K \rightarrow 2\pi$ amplitude at $s\simeq 0.96m_\pi^2$ which is
prettty near to the Cabbibo and Gell-Mann point, and the latter at $s\simeq 0.92m_\pi^2$ which
is a larger violation of this theorem. The only point where there is no violation of the Cabibbo
and Gell-Mann theorem is $s_0=m_\pi^2$ which is totally expected.

 Normalising the factor $\sqrt{2}Cf_\pi$ to be unity, for $s_0=0, m_\pi^2$, and $2m_\pi^2$ we
have, respectively, the absolute value of the physical matrix element ($s=m_K^2$) to be 19, 18.2
and 15.4 which shows some sensitivity on the choice of $s_0$. Fortunately for our problem,
$s_0=m_\pi^2$ as required by the Cabibbo and Gell-Mann theorem.

As $s_0$ approaches the branch point,  the violation of this theorem becomes larger. For example
at 
$s_0=3m_\pi^2$, the zero of the matrix element is at $0.5m_\pi^2$ which is a large violation.
This is due to the square root threshold singularity of the matrix element, resulted by the
threshold behavior of the S-wave $\pi\pi$ phase shift, $\delta \rightarrow 0$ as
$\sqrt{s-4m_\pi^2}$.

In Fig. (1) the function $\Omega(s,0)$ derived in
\cite{truong1, truong4, truong6, truong5} is plotted as a function of s.  For $s<4m_\pi^2$ this
function is real. For $s>4m_\pi^2$, $\Omega(s,0)$ is complex, only its real part is plotted. As
one can see at $s=4 m_\pi^2$ there is a cusp associated with what was known as the square root
(threshold) singularity, but is now misnamed  as the "chiral logarithm" singularity
\cite{truong3, truong5}. This singularity is due to the threshold behaviour of the phase shift as
discussed above. On the same graph, the imaginary part of
$\Omega(s,0)$ is plotted as a function of s.

As can be seen in Fig. (1), because of the square root singularity, a power series expansion
for  $\Omega(s)$  near to the branch point requires many terms to give
the correct energy dependence;  the series converges inside a  small circle with
the center at $s_0$ and with a radius $\mid4m_\pi^2-s_0\mid$.

Our approach to this problem is heavily based on the reinterpretation of the tree CL as a
power series expansion below the $2\pi$ threshold even in the presence of the strong $\pi \pi$
(final state) interaction and on the use of the current algebra low energy theorems. It is  quite
different from the spirit of the reference
\cite{buras} where ChPT is used and hence the assumption on the derivative on $s_0=m_\pi^2$ has to
be made. Our reinterpretation of the tree CL leads naturally to this condition. Some of the points
discussed in their article are clarified in this article.

This author would like to thank Luis Oliver for pointing out the existence of the reference
\cite{buras}.

\newpage

{\bf Figure Captions}

Fig. 1~:The real part of the function $\Omega(s,0)$ as a function of $s$ (in the unit
$m_\pi^2=1$) is shown by the solid line; the imaginary part of $\Omega(s,0)$ is shown by the
dotted line.

\newpage
\begin{figure}
\epsfbox{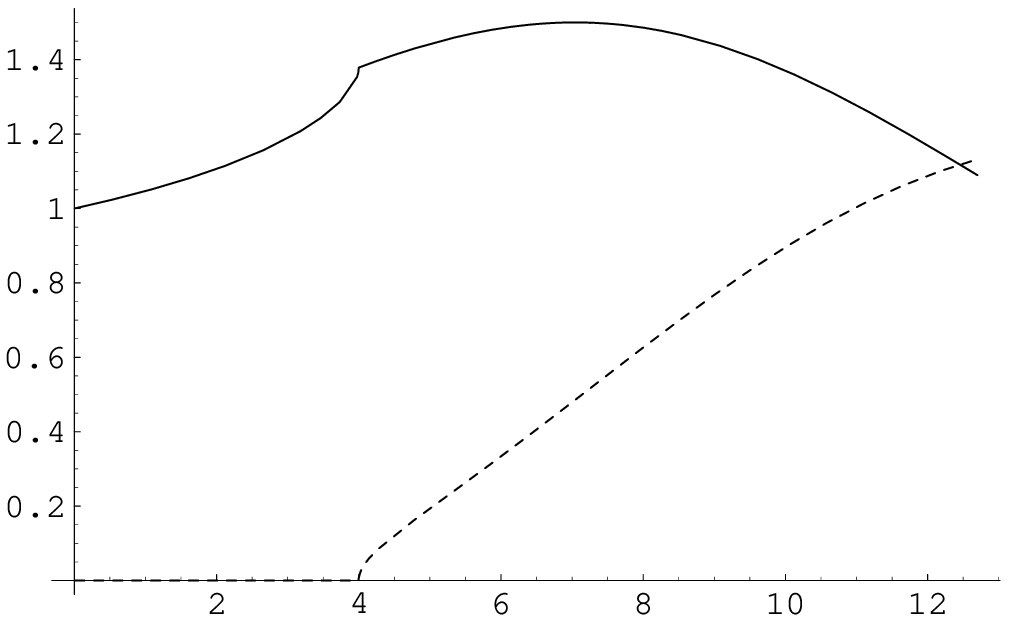}
\caption{}
\label{Fig. 1}
\end{figure}


\newpage
\begin{thebibliography}{99}
\bibitem{truong1} T. N. Truong, Phys. Lett. {\bf B207} (1988) 495.
\bibitem{pallante} E. Pallante and A. Pich, hep-ph/9911233.

\bibitem{neubert} M. Neubert and B. Stech, Phys. Rev. {\bf D44} (1991) 775 .
\bibitem{buras} A. J. Buras, M. Ciuchini, E. Franco, G. Isidori, G. Martinelli and L. Silvestrini,
hep-ph/0002116.
\bibitem{adler} For a review, see \emph{Current Algebra} by S. L. Adler and R. Dashen, W.A.
Benjamin, New York 1968.
\bibitem{cab} N. Cabibbo, Phys. Rev. Lett. {\bf12} (1964)  62; M. Gell-Mann, Phys. Rev. Lett.
{\bf 12} (1964) 155. See also more recent references cited in \cite{buras}.

\bibitem{truong2} T. N. Truong, Phys. Lett. {\bf B313} (1993)  221.
\bibitem{truong3} T. N. Truong, Phys. Lett. {\bf B99} (1981) 154 .
\bibitem{truong4} C. Roiesnel and T. N. Truong, Nucl. Phys. {\bf B187} (1981) 293. 
\bibitem{truong6}  T. N.
Truong, Acta Phys. Pol. {\bf B15} (1984)  633; T. N. Truong, in \emph{Wandering in the
Fields}, edited by K. Kawarabayashi and A. Ukawa (World Scientific, Singapore, 1987).
\bibitem{truong5}  T. N. Truong, Phys. Rev. Lett. {\bf 61}  (1988) 2526.
\bibitem{omnes} R. Omnes, Nuovo Cimento {\bf 8}  (1958) 316.
\bibitem{mus} N. I. Muskhelishvili, \emph{ Singular Integral Equations} (Noordhoof,
Groningen, 1953).
\bibitem{cronin} J. A. Cronin, Phys. Rev. {\bf 161} (1967) 1483 . For a more up to date list
references on this subject, see references given in \cite{buras}. 
\bibitem{all} For a complete list of references on the lattice gauge theory calculations and
approximate schemes, see references listed in \cite{buras}.



\end{thebibliography}
\end{document}